# Estimating the Tail Index by using Model Averaging

J. Martin van Zyl

*Department of Mathematical Statistics and Actuarial Science, University of the Free State, Bloemfontein, South Africa*

E-mail: wwjvz@ufs.ac.za

Abstract: The ideas of model averaging are used to find weights in peak-over-threshold problems using a possible range of thresholds. A range of the largest observations are chosen and considered as possible thresholds, each time performing estimation. Weights based on an information criterion for each threshold are calculated. A weighted estimate of the threshold and shape parameter can be calculated.

## 1. Introduction

Often there are various models which all performs well and each model has some properties which may be better than others with respect to certain aspects of the problem. This problem occurs mostly in problems with one sample and various models with a different number of parameters. In such problems model averaging can be performed. An overview of model averaging is given by Moral-Benito (2013). There is a Bayesian approach called Bayesian model averaging (BMA) and the frequentist model averaging (FMA) approach. In this work weights are proposed for different thresholds in peak over threshold (POT) problems as introduced by Pickands (1975). Thus rather than using a specific point, many points are taken into consideration when performing estimation. The weights are used to calculate a weighted average over a range of estimated indexes of a Pareto distribution. A range of the largest observations in a sample is considered as possible thresholds, when performing estimation. The weights must be comparable over different sample sizes and the average likelihood for a specific threshold is used to calculate the weight for that specific estimate. Of course model averaging can also be performed using various estimation methods and the same threshold, but that is not considered in this work.

The weight which will be used are of the form of those proposed by Buckland *et al.* (1997), that is the weight for model *r*, where there are *q* possible models, using an information criterion for say model $r$, $I_r$, is:

$$w_r = \exp(-I_r/2) / \sum_{r=1}^{q} \exp(-I_r/2), \quad (1)$$

$$I_r = -2\log(L_r) + \varphi_r .$$

$L_r$ is the maximized likelihood for model *r*, $\varphi_r$ is a penalty term, involving the number of parameters used in the model. It can be shown that the average log-likelihood which cancels the effect of sample size, is also valid. For a model with *p* parameters where a sample of size *n* is available, the two most well-known criteria are the Akaike Information Criterion (AIC) with $\varphi_r = 2p$ (Akaike, 1973). The Bayesian Information Criterion (BIC) has a penalty factor $\varphi_r = p\log(n)$. BIC is also often called SBIC or Schwarz BIC named after the author who wrote a paper on estimating the dimension of a model and introduced BIC (Schwarz, 1976).

The model selection work was derived to be used on one set of data where there are possible models and to either select one model or use model averaging. The information criteria and also specifically the likelihood is a function of the sample size. It can be shown that for example the log-likelihood of a Pareto distribution is a linearly increasing function of the sample size, so this must be taken into account when comparing models with different sample sizes.

Consider the original derivation of the AIC. Suppose that a possible model assumes the density of the data is $p(y,\theta)$. A sample, $y_1,...,y_n$, with unknown density $f(y;\theta)$ is available. Let $\hat{\theta}$ denote the MLE of $\theta$ assuming that the observations have density $p$. The distance or quality of $\hat{p}(y;\hat{\theta})$ as an estimate of $f(y;\theta)$ can be estimated using the Kullbach-Leibner distance

$$K(f,\hat{p}) = \int f(y;\theta)\log(f(y;\theta)/\hat{p}(y;\hat{\theta}))dy$$

$$= \int f(y;\theta)\log(f(y;\theta))dy - \int f(y;\theta)\log(\hat{p}(y;\hat{\theta}))dy \ .$$

The first term is fixed, thus minimizing $K(f,\hat{p})$ is equivalent to maximizing the second term. An estimate of the second term is

$$\frac{1}{n}\sum_{j=1}^{n}\log(p(y_j;\hat{\theta})) = l_j(\hat{\theta})/n \,,$$

where $l_j(\hat{\theta}) = l(y_1,...,y_n;\hat{\theta})$ denote the log-likelihood assuming density $p$ . This is a biased estimate of the true integral, and it was shown that the bias is approximately the dimension $d$ of $\theta$, divided by $n$ . Using the approximation it follows that the information criterion is $\hat{K} = l(\hat{\theta})/n - d/n$ . This expression was later multiplied by the constant, $-2n$ , and this expression minimized. Let

$$I_r = l_r(\hat{\theta})/n_r - d/n_r, r = 1,...,q \ . \qquad (2)$$

The proposed weights when $q$ thresholds are considered will then be of the form:

$$w_j = \exp(I_j / 2) / \sum_{r=1}^{q} \exp(I_r / 2),$$

where $n_r$ observations are used for the r-th possible threshold when estimating the tail index and the log-likelihood in the point $\hat{\theta} = \hat{\theta}_{MLE}$ for this threshold is $l_r(\hat{\theta})$.

## 2. Weights for two approaches to estimate the tail index

Two of the approaches used to estimate the tail index is by using the Pareto assumption and assuming observations above a threshold has a Pareto distribution or a generalized Pareto distribution (GPD), and the other makes use of the power tail form of the survival function above the threshold and performing regression. If one rewrite a simple of estimators it can be seen as a type of weighted estimate with the largest order statistics having the most weight, since the largest observation will every time be included. In a sample of size $n$, if the largest $k_1$ to $k_2$ observations are considered as possible thresholds, to calculate a weighted average, every order statistic as a possible threshold need not be included, and for example every tenth order statistic can also be used to calculate the average.

## 2.1 The Pareto assumption

In this section the likelihood assuming a Pareto distribution above a threshold will be considered. It was found that when using the GPD distribution of excesses above a threshold, when many points are considered there are almost always a few numerical convergence problems at certain sample sizes in POT problems. In the ideal circumstances when estimating the parameters of GPD distributed data these convergence problems occur almost never, but in POT problems it is a factor. Theoretically in very large samples the GPD with should perform better and in such samples if the threshold is very large and there are many observations above the threshold few numerical problems would be encountered.

In the usual practical problems sample sizes are often a few thousand and it was found that the Pareto assumption using the weighted model performs well and is from a numerical and practical viewpoint a better method to use. The pure Pareto assumption is easy to calculate, there are no numerical problems and performs good. So in the simulation study only the pure Pareto assumption was considered. The Hill estimator can also be derived by assuming that the largest observations above a threshold are Pareto distribution.

If a Pareto distribution is assumed above a threshold, similar type of weights can be derived. Suppose a sample of size $n, x_1, ...., x_n$, is available. The corresponding order statistics are $x_{(1)} \le x_{(2)} \le ... \le x_{(n)}$. Consider the Pareto distribution in general. Let

$$p(x;\alpha,\beta) = \alpha\beta^{\alpha} x^{-(\alpha+1)}, \; x \ge \beta \,. \qquad (3)$$

The log-likelihood can be written as

$$l(x_1,...,x_n;\alpha,\beta) = n\log(\alpha) - n\log(\beta) - (\alpha+1)\sum_{j=1}^{n}\log(x_j/\beta)\,.$$

The ML estimates of the parameters are

$$\hat{\beta} = x_{(1)},\ \hat{\alpha} = n/\sum_{j=1}^{n}\log(x_j/\hat{\beta})\,,$$

leading to:

$$l(x_1,...,x_n;\hat{\alpha},\hat{\beta})/n = \log(\hat{\alpha}) - \log(x_{(1)}) - (\hat{\alpha}+1)/\hat{\alpha}\,. \qquad (4)$$

If the $m$ largest observations above a threshold in a POT problem are considered, it follows that:

$$\hat{\beta}_m = x_{(n-m)},\ \alpha_m = m/\sum_{r=1}^{m}\log(x_{(n-m+r)}/\hat{\beta}_m)\,, \qquad (5)$$

$$l(x_{(n-m+1)},...,x_{(n)};\hat{\alpha}_m,\hat{\beta}_m)/m = \log(\hat{\alpha}_m) - \log(x_{(n-m)}) - (\hat{\alpha}_m + 1)/\hat{\alpha}_m,$$

$$I_{n-m} = l(x_{(n-m+1)},...,x_{(n)};\hat{\alpha}_m,\hat{\beta}_m)/m - 2/m. \qquad (6)$$

In a sample of size $n$, if the largest $k_1$ to $k_2$ observations are considered as possible thresholds, the equations are used to calculate the weights for each $m = n - k_1,...,n - k_2$.

$$w_m = \exp(I_m/2) / \sum_{j=n-k_1}^{n-k_2} \exp(I_j/2),$$

$w_m$ denote the weight if the largest *m* observations are used and the threshold is assumed to be $x_{(n-m)}$.

A more complex assumption would be to make use of the peak over threshold theorem where it was shown that above a certain threshold, say u, the excesses above the threshold are generalized Pareto distributed. Let $Y = X - u$, then it follows that:

$$F_u(y) = 1 - (1 + (\xi / \sigma) y)^{-1/\xi},$$

where $\xi$ denotes the shape parameter and the index is, $\alpha = 1/\xi$, $\sigma$ a scale parameter.

The log-likelihood for a given threshold, $u = x_{(n-k)}$, using as sample the $k$ excesses above the threshold evaluated at the ML estimators is;

$$l(x_1, ..., x_n, u; \sigma, \xi) = -k \log(\hat{\sigma}) - (1/\hat{\xi} + 1) \sum_{j=n-k+1}^{n} \log(1 + (\hat{\xi} / \hat{\sigma})(x_j - u)).$$

.

The information criterion to calculate the weights is

$$I_k = l(x_1, ..., x_n, u; \hat{\sigma}, \hat{\xi}) / k - 2/k .$$

This assumption works excellent if the ideal assumptions are met, but in POT problems and smaller sample sizes and then fitting it to data above a threshold often lead to convergence problems and the Pareto assumption was found to be a better practical method, when using the weighted estimation.

## 2.2 Power tail assumption

If it is assumed that the largest observations from a distribution $P$ obeys the power law and that $1-P(x)=cx^{-\alpha}$ or

$$\log(1-P(x))=\log(c)-\alpha\log(x), \qquad (7)$$

and the empirical distribution is used to estimate $P(x)$, linear regression can be performed to estimate the tail index. If normal error terms are assumed the log-likelihood is of the form, the *m* largest observations are used, ignoring constants,

$$l(\hat{\alpha}_m)/m \propto -\log(\hat{\sigma}_m)\ , \quad \hat{\sigma}_m^2=\frac{1}{m}\sum_{r=1}^{m}\hat{\varepsilon}^2\ , \qquad (8)$$

where $\hat{\varepsilon}_1,...,\hat{\varepsilon}_m$, denote the residuals when using the largest *m* observations to estimate $\alpha$. Let

$$I_m = -\log(\hat{\sigma}_m) \text{ -2/m,} \qquad (9)$$

then it follows that

$$w_m = \exp(I_m / 2) / \sum_{j=n-k_1}^{n-k_2} \exp(I_j / 2),$$

for possible thresholds $x_{(n-k_1)},..., x_{(n-k_2)}$.

## 2. Simulation Study and an Application

In this section POT estimation will be performed on data simulated from the stable distribution, a t-distribution and also a GPD and the largest observations in each sample used to estimate the index of the tails and also a weighted estimate of the threshold. For the stable and t-distribution data will be simulated from symmet-ric distributions and the estimation performed on the absolute values. For all the distribution the shape parameter is denoted by $\xi$, with index $\alpha = 1/\xi$. The skewness, location parameter and scale parameters are denoted by $\beta, \mu, \sigma$ for the stable distributions. For the t-distribution the degrees of freedom denoted by $\nu$ with location and scale parameter $\mu$ and $\sigma$. The GPD has location parameter $\mu$, and scale parameter $\sigma$. The mean square error (MSE) and bias are given, based on using the largest 50 to 500 possible thresholds, and $m = 5000$ samples generated of size $n = 2500$ each.

Note that in all simulations the results are given as the bias and MSE with respect to the index $\alpha$.

| | | Pareto likelihood | | | Power-tail regression | | |
|---|---|---|---|---|---|---|---|
| | | Estimated threshold | MSE | bias | Estimated threshold | MSE | bias |
| Stable distribution | $\alpha = 0.8$ | 9.3521 | 0.0183 | 0.0018 | 13.9264 | 0.0343 | 0.0056 |
| | $\alpha = 1$ | 5.7371 | 0.0048 | 0.0026 | 7.1862 | 0.0362 | 0.0087 |
| $\beta = 0, \mu = 0, \sigma = 1$ | $\alpha = 1.5$ | 2.9541 | -0.2291 | 0.0614 | 3.1972 | -0.0755 | 0.0289 |
| | $\alpha = 1.8$ | 2.4516 | -0.9173 | 0.8724 | 2.5426 | -0.5367 | 0.3818 |
| t-distribution $\mu = 0, \sigma = 1$ | $\nu = 3$ | 2.3540 | 0.5760 | 0.3478 | 2.3874 | 0.4533 | 0.2532 |
| | $\nu = 5$ | 2.0338 | 1.8601 | 3.4861 | 2.0630 | 1.5418 | 2.4491 |
| GPD $\mu = 1, \sigma = 1$ | $\xi = 0.25$ | 3.1341 | 1.8752 | 3.5272 | 3.2947 | 1.6205 | 2.6581 |
| | $\xi = 0.5$ | 4.2377 | 0.4673 | 0.2245 | 4.5822 | 0.3725 | 0.1570 |
| | $\xi = 0.75$ | 5.8282 | 0.1620 | 0.0295 | 6.7127 | 0.1311 | 0.0267 |
| | $\xi = 1$ | 8.1288 | 0.0645 | 0.0064 | 10.4239 | 0.0613 | 0.0105 |
| | $\xi = 1.2$ | 10.6333 | 0.0311 | 0.0025 | 15.2669 | 0.0413 | 0.0062 |

Table 1. Estimation of the shape parameter using the largest observations in sample of size n=2500. MSE and bias given.

It can be seen that the success of the estimation is sensitive to how heavy-tailed the data is and good results found especially in very heavy-tailed data. A sample of size $n = 2500$ is relatively small for POT problems, but especially where $\alpha < 2$ ($\xi > 0.5$), both methods give good estimates of the shape parameter. The estimated threshold is smaller than the one using regression.

In table 2 the scale parameter $\sigma = 2$ was used, the mean square error (MSE) and bias are given, based on using the largest 50 to 500 possible thresholds, and $m = 5000$ of size $n = 5000$ each.

| | | Pareto likelihood | | | Power-tail regression | | |
|---|---|---|---|---|---|---|---|
| | | Estimated threshold | MSE | bias | Estimated threshold | MSE | bias |
| Stable distribution $\beta = 0, \mu = 0, \sigma = 2$ | $\alpha = 0.8$ | 45.5218 | 0.0052 | 0.0018 | 67.2706 | 0.0328 | 0.0061 |
| | $\alpha = 1$ | 23.1191 | 0.0014 | 0.0026 | 28.9841 | 0.0364 | 0.0087 |
| | $\alpha = 1.5$ | 8.6752 | -0.1684 | 0.0371 | 9.4345 | -0.0206 | 0.0216 |
| | $\alpha = 1.8$ | 6.1844 | -0.8881 | 0.8181 | 6.6113 | -0.3526 | 0.1954 |
| t-distribution $\mu = 0, \sigma = 2$ | $\nu = 3$ | 6.3510 | 0.3594 | 0.1494 | 6.3677 | 0.3255 | 0.1586 |
| | $\nu = 5$ | 5.1704 | 1.4053 | 2.0091 | 5.1637 | 1.2074 | 1.5507 |
| GPD $\mu = 1, \sigma = 2$ | $\xi = 0.25$ | 8.9674 | 1.5637 | 2.4597 | 9.1706 | 1.3771 | 1.9367 |
| | $\xi = 0.5$ | 13.6430 | 0.3195 | 0.1104 | 14.3228 | 0.2795 | 0.1002 |
| | $\xi = 0.75$ | 21.3435 | 0.0960 | 0.0131 | 24.0717 | 0.1040 | 0.0221 |
| | $\xi = 1$ | 34.3364 | 0.0307 | 0.0034 | 43.5135 | 0.0539 | 0.0101 |
| | $\xi = 1.2$ | 51.0214 | 0.0099 | 0.0018 | 72.1831 | 0.0329 | 0.0061 |

Table 2. Estimation of the shape parameter using the largest observations in sample of size n=5000. MSE and bias given.

It can be seen that the success of the estimation is again very sensitive to how heavy-tailed the data is and good results found especially in very heavy-tailed data. especially where $\alpha < 2$ ($\xi > 0.5$), both methods give good estimates of the shape parameter. The estimated thresholds are much larger than in the $n = 2500$ samples, which may be an indication that the distributional behaviour of the largest observations is not yet pure Pareto in the smaller sample.

The bias of the estimated parameters are better in the larger sample as can be expected. For example to estimate $\alpha$ of a Cauchy distribution, that is $\alpha = 1, \beta = 0$ for a stable distribution using the Pareto assumption results in a bias of 0.0048 in a sample of size $n = 2500$ and 0.0014 in a sample of size $n = 5000$, with similar estimated MSE's of 0.0026.

In figure 1 a histogram is shown with 5000 estimated parameters using the Pareto assumption and estimation is based on the largest order statistics as a possible threshold, from the 500 largest to the 50 largest. The samples of size $n = 5000$ are form a stable distribution with $\alpha = 1.2, \sigma = 2, \beta = 0$ . The mean of the 5000 estimated indices is 1.1709, with MSE 0.0119.

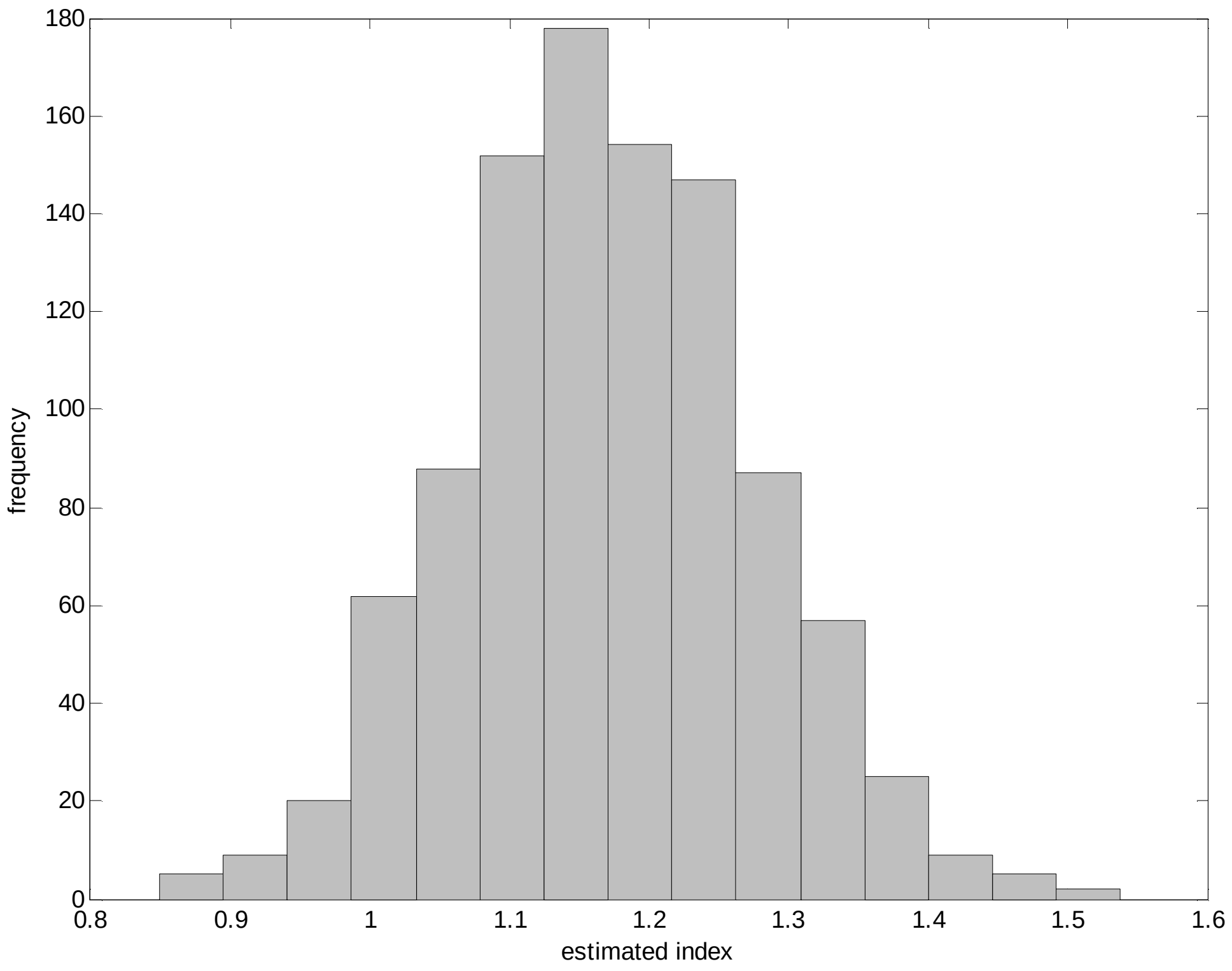

Figure 1. Histogram of 5000 estimated indices in POT problems in samples of size n=5000 each. Samples stable distributed with index $\alpha = 1.2, \sigma = 2, \beta = 0$ .

# 3. Application to the Danish Fire Losses

This set of data was analysed by various researchers. Some of the references are McNeil (1997, 1999), Resnick (1997), Lee et al. (2012), Embrechts et al. (1997), Zivot and Wang (2003). There are 2156 losses larger than 1 million Danish kroner. The total sample size is 2492 observations.

When using different assumptions and models different estimates of the threshold were found. Zivot and Wang (2003) estimated the threshold as 5.28 (millions) using the Hill quantile estimator and the ML GPD estimator is 5.20. Stable estimates of the index using the Hill estimator and a threshold in the region of 10 was found to be approximately in the region of 2.01. The estimates using the weighted approach are as follows:

- Using the weighted average assuming the Pareto model, a threshold is estimated as 4.7154 , which

means the largest 276 observations are used. The index is estimated as $\hat{\alpha} = 1.4435\,(\hat{\xi} = 0.6928)$ .

- If the power tail regression approach is follows the threshold is estimated as 5.3061, which means the largest 234 observations are included. The index is estimated as $\hat{\alpha} = 1.4521\,(\hat{\xi} = 0.6887)$ .

In figures 2 and 3 model using the regression approach is shown versus the actual values. It can be seen that the fit is good, especially concerning the largest observations in the sample.

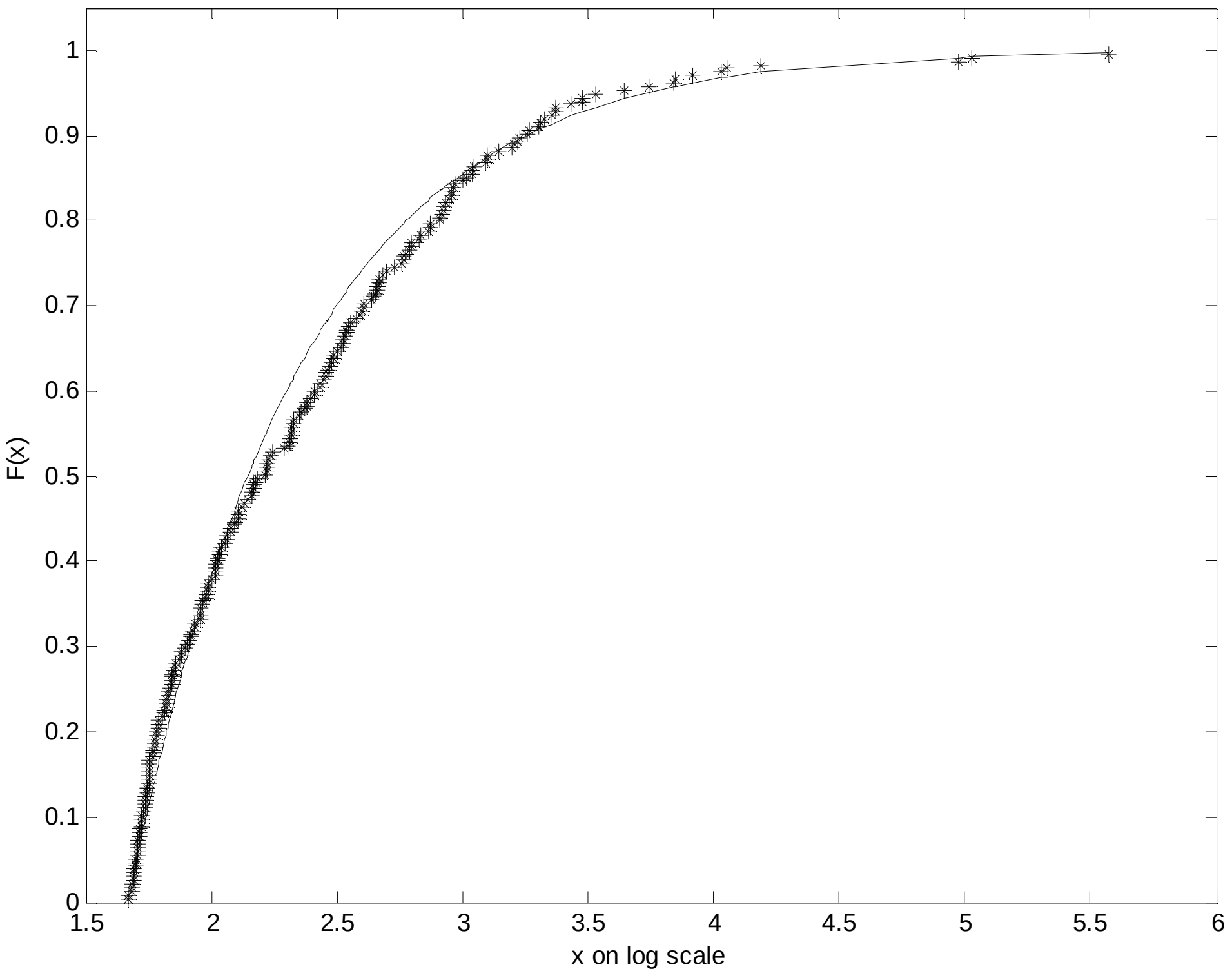

Figure 2. Fitted Pareto distribution to values over estimated threshold on a log scale.

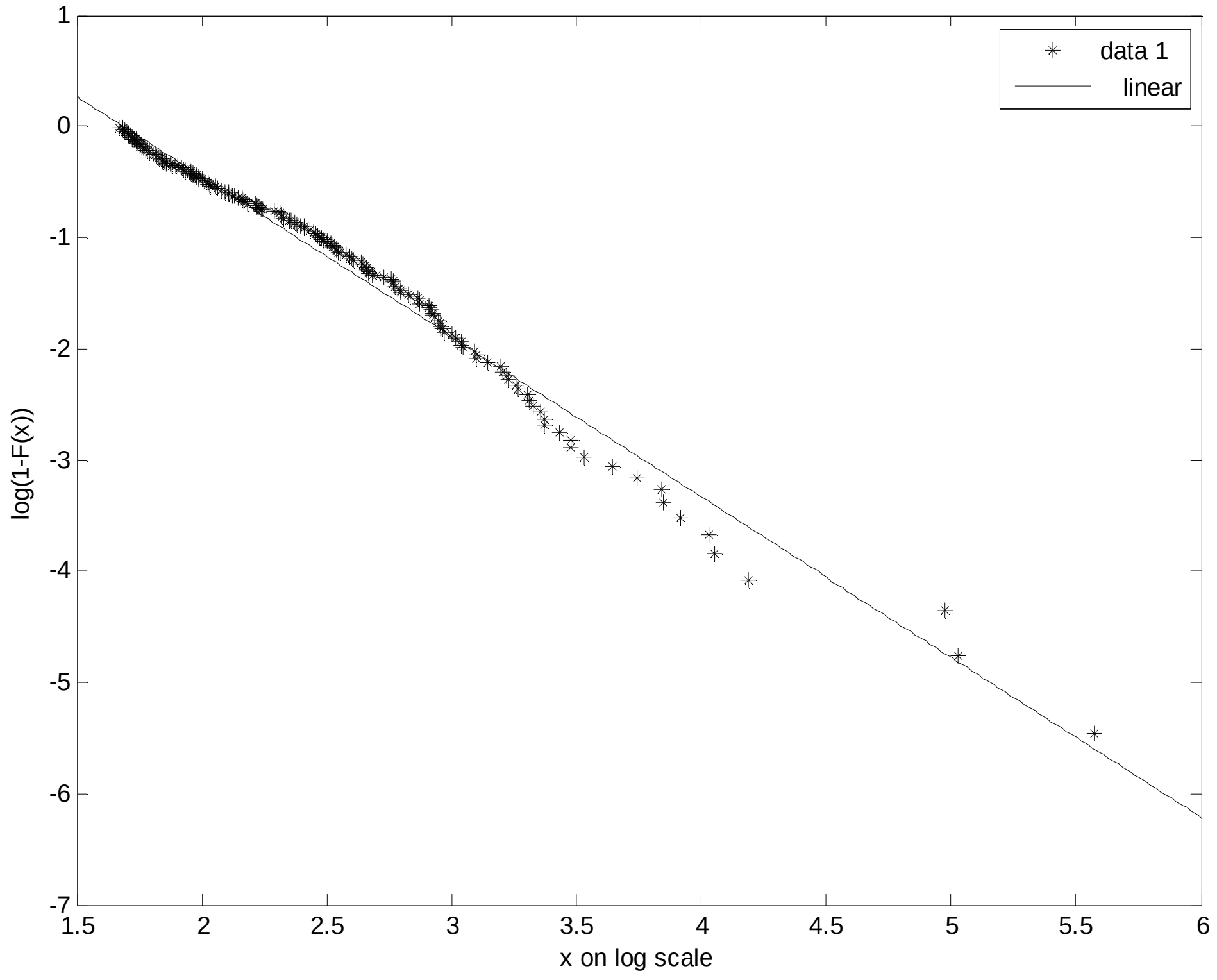

Figure 3. Quabtile plot of estimated distribution on a log-scale

In the weighted method the largest values will be taken into account when calculating each weight, and the process is inherently so that the largest observations are considered more important. It can be seen in the figures that the estimated distribution fits especially the largest observations good.

## 4. Conclusions

In this work the very basic ideas of model averaging is applied to estimate the parameters of heavy-tailed distributions, using in POT problems. It was found to perform good. One advantage of this procedure is, that a reasonable estimate of the threshold can be found without having to look at figures. This estimate is dependent on the quality of the data and will be good if the sample is such that the POT technique can be applied.

When checking the performance of estimation techniques using simulated samples in POT problems, one cannot study charts each time, and such a procedure might be of help in such simulation studies giving a reasonable estimate of the threshold.